# Representation of Boolean functions in terms of quantum computation

Yu.I. Bogdanov*[abc], N.A. Bogdanova[ab], D.V. Fastovets**[ab], V.F. Lukichev[a]

[a]Valiev Institute of Physics and Techonology of Russian Academy of Sciences, Russia, Moscow; [b]National Research University of Electronic Technology (MIET), Russia, Moscow; [c]National Research Nuclear University (MEPhI), Russia, Moscow

## ABSTRACT

The relationship between quantum physics and discrete mathematics is reviewed in this article. The Boolean functions unitary representation is considered. The relationship between Zhegalkin polynomial, which defines the algebraic normal form of Boolean function, and quantum logic circuits is described. It is shown that quantum information approach provides simple algorithm to construct Zhegalkin polynomial using truth table. Developed methods and algorithms have arbitrary Boolean function generalization with multibit input and multibit output. Such generalization allows us to use many-valued logic ($k$-valued logic, where $k$ is a prime number). Developed methods and algorithms can significantly improve quantum technology realization. The presented approach is the baseline for transition from classical machine logic to quantum hardware.

## 1. INTRODUCTION

Discrete mathematics is an important area of mathematic science, which explores the properties of different discrete objects: graphs [1], Boolean functions [2-4], finite-state machines and etc. The methods of discrete mathematics have important application in various scientific fields, such as logic elements realization of electronic devices, information security [5], transport links optimization, business models construction and etc.

The discrete systems have been explored in quantum mechanics and quantum information theory. The discretization and quantization have a similar significance. But, for a long time, discrete mathematics had developed without in-touch with quantum theory.

The Zhegalkin polynomial [6] is an important object of discrete mathematics which has important application in quantum circuits design. The set of all quantum circuits (with X gate and its condition analogues) can be constructed using injective function to the set of all Zhegalkin polynomials. In other words, arbitrary Zhegalkin polynomial can be transformed to quantum circuit. It is described below in our paper. An important feature is the simple construction of the scheme. We have demonstrated an effective method to Zhegalkin polynomial constructing using the truth table of the original function.

Gates X, CNOT, CCNOT (and etc.) availability in the circuits constructed from Zhegalkin polynomials is explained by the fact that there are similar transforms in classical logic [7]. Quantum mechanics provides resources in the unitary operations form. Such operations allow us not only to construct quantum analogues of classical circuits, but to generalize them [8-9]. For example, by introducing the Boolean function inverse transform, and considering quantum superpositions of basis states. Thus, quantum Boolean functions are better objects in comparison with classical binary functions. The quantum Boolean functions are the subset of quantum transform general class.

The developed approach provides the construction of important tools for quantum information processing methods. For example, for quantum oracle transform realization in different algorithms.

*bogdanov_yurii@inbox.ru; **fast93@mail.ru

The modern information society infrastructure is based on the Boolean algebra principles and the methods of discrete mathematics. From the middle of 20th century to the present days, information technology is the main global economy driver.

In our days, it is becoming increasingly obvious that quantum information technologies should be the driver of information technologies development in the coming years and decades. Thus, it is necessary to integrate the methods of discrete mathematics and quantum information technology.

## 2. CONSTRUCTION OF A UNITARY TRANSFORMATION CORRESPONDING TO A GIVEN BOOLEAN FUNCTION

According to quantum information technology [10,11] the quantum realization of Boolean function $f(x)$ is a transformation of two-particle state $|x, y\rangle$ to $|x, y \oplus f(x)\rangle$:

$$|x,y\rangle \xrightarrow{f} |x, y \oplus f(x)\rangle. \qquad (1)$$

Here, $x$ the state of $n$-qubit register (function's input), $y$ - output register (one- or multi-qubit). Symbol $\oplus$ means addition mod 2. The graphical interpretation of (1) is shown on Fig. 1.

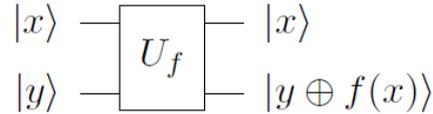

Figure 1. Quantum circuit for Boolean function $f(x)$ computation.

Note that, by default, it is usually assumed that there is no source register $x$ on the circuit's input and output in discrete mathematics manuals. In this case, formula (1) is transformed to the following form: $|x\rangle \xrightarrow{f} |f(x)\rangle$. The use of a more general definition in quantum computing makes it possible to provide unitary invertible character of computations. In this section, we assume that the output register $y$ has one qubit. In this case $U_f$ - $2^{n+1} \times 2^{n+1}$ unitary matrix.

Let us consider the simplest case: the function $f(x)$ has a one input bit and a one output bit. There are four such functions (Table 1).

Table 1. One-bit functions truth table.

| $x$ | $f_0$ | $f_1$ | $f_2$ | $f_3$ |
|---|---|---|---|---|
| 0 | 0 | 0 | 1 | 1 |
| 1 | 0 | 1 | 0 | 1 |

Note that, $f_0 = 0$ and $f_3 = 1$ - constant functions, $f_1 = x$ and $f_2 = x \oplus 1$ - variable functions. Let $I = \begin{pmatrix} 1 & 0 \\ 0 & 1 \end{pmatrix}$ - identity matrix defining the identity transform, and $X = \begin{pmatrix} 0 & 1 \\ 1 & 0 \end{pmatrix}$ - matrix defining the inversion (NOT transform). Based on function definition (1), it is easy to show that the unitary transforms matrices corresponding to the four Boolean functions in Table 1 are

$$U_0 = \begin{pmatrix} I & 0 \\ 0 & I \end{pmatrix}, U_1 = \begin{pmatrix} I & 0 \\ 0 & X \end{pmatrix}, U_2 = \begin{pmatrix} X & 0 \\ 0 & I \end{pmatrix}, U_3 = \begin{pmatrix} X & 0 \\ 0 & X \end{pmatrix}. \quad (2)$$

All of this matrices are $4 \times 4$ and have a block-diagonal view. To avoid misdirection, we note that zero in formula (2) is a $2 \times 2$ matrix zero: $0 = \begin{pmatrix} 0 & 0 \\ 0 & 0 \end{pmatrix}$. The construction principle of matrices (2) is very simple: the zero in truth table is matched to the matrix $I$, and one matched to $X$.

The same construction principle holds true for multi-bit Boolean functions. In this case, the transform matrix $U_f$ for $n$-bit Boolean function is composed of $2^n$ blocks: $I$ or $X$ only. Let us formulate this statement in the form of the following general statement.

<u>Statement 1</u> (on the block-diagonal character of Boolean transforms). The unitary matrix corresponding to a Boolean function has a block-diagonal form. Function value $f(x) = 0$ corresponds to the $I$ matrix, and function value $f(x) = 1$ corresponds to the $X$ matrix. The $n$-bit argument $x$ values are ordered in ascending order from $0$ to $2^n - 1$.

Statement 1 implies that there are $2^{(2^n)}$ block-diagonal matrices and the same number of Boolean functions.

$n = 1$: $2^2 = 4$ Boolean functions, $U_f$ - $4 \times 4$ matrices;

$n = 2$: $2^{(2^2)} = 16$ Boolean functions, $U_f$ - $8 \times 8$ matrices;

$n = 3$: $2^{(2^3)} = 256$ Boolean functions, $U_f$ - $16 \times 16$ matrices;

It is convenient to represent quantum transforms in terms of graphical circuits. Quantum circuits for one-bit Boolean functions (Table 1) are shown in Fig. 2.

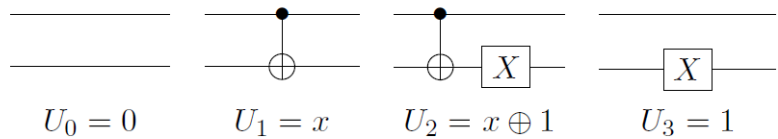

Figure 2. Quantum circuits for one-bit Boolean functions.

The function $f_0 = 0$ corresponding to the identical transform are described by two quantum wires without any gates. The function $f_3 = 1$ corresponds to the gate $X$ (NOT) action on the lower qubit (output register). The function $f_1 = x$ corresponds to important quantum logical gate CNOT (controlled NOT). According to the truth table (Table 1), the CNOT gate acts as follows: CNOT changes the state of the second (controlled, lower in the Fig. 2) qubit, if the first (controlling, upper in the Fig. 2) qubit is in the state $|1\rangle$. Finally, the function $f_2 = x \oplus 1$ corresponds to the sequential action of the operators CNOT and $X$ (NOT).

The composition (addition mod 2) of two Boolean functions (according to Statement 1) is the multiplication of two block-diagonal unitary matrices. The matrix identity $I \cdot I = I^2 = I$ corresponds to the Boolean identity $0 + 0 = 0$. The matrix ratio $I \cdot X = X \cdot I = X$ corresponds to the Boolean identity $0 + 1 = 1$. Finally, the matrix ratio $X \cdot X = X^2 = I$ corresponds to the Boolean identity $1 + 1 = 0$. Here and throughout, we mean the addition mod 2 under the sum sign.

Let one-bit input function $x$ be vector $x = \begin{pmatrix} 0 \\ 1 \end{pmatrix}$. We will consider this vector as a basis vector of a two-dimensional vector space: $e_1 = x = \begin{pmatrix} 0 \\ 1 \end{pmatrix}$. Then $e_0 = x + 1 = \begin{pmatrix} 1 \\ 0 \end{pmatrix}$ - is the second basis vector. It is easy to verify the following identities:

$$e_0 + e_1 = 1 = \begin{pmatrix} 1 \\ 1 \end{pmatrix}, e_0 + e_0 = 0 = \begin{pmatrix} 0 \\ 0 \end{pmatrix}, e_1 + e_1 = 0 = \begin{pmatrix} 0 \\ 0 \end{pmatrix}.$$

The logical basis function $e_0$ corresponds to the polynomial $1 + x = 1 \cdot x^0 + 1 \cdot x^1$. Coefficients of this polynomial form a column vector $p_0 = \begin{pmatrix} 1 \\ 1 \end{pmatrix}$. Here, $x^0 = \begin{pmatrix} 1 \\ 1 \end{pmatrix}, x^1 = \begin{pmatrix} 0 \\ 1 \end{pmatrix}$ - are zero and first degrees of vector $x = \begin{pmatrix} 0 \\ 1 \end{pmatrix}$. The logical basis function $e_1$ corresponds to the polynomial $x = 0 \cdot x^0 + 1 \cdot x^1$. Coefficients form a column vector $p_0 = \begin{pmatrix} 0 \\ 1 \end{pmatrix}$. The basis function equation $e_0 = 1 + x = 1 \cdot x^0 + 1 \cdot x^1 = \begin{pmatrix} 1 \\ 0 \end{pmatrix}$ corresponds to the column-function $\begin{pmatrix} 1 \\ 0 \end{pmatrix}$, corresponding to the polynomial $1 \cdot x^0 + 1 \cdot x^1$. The equation $e_0 = x + 1$ can be interpreted as numeric function $e_0 = 1$ at $x = 0$ and $e_0 = 0$ at $x = 1$. The vector representation corresponds to the logical function as a single object. It is a very useful approach. Similar considerations are valid for the basis logical function $e_1 = x = 0 \cdot x^0 + 1 \cdot x^1 = \begin{pmatrix} 0 \\ 1 \end{pmatrix}$. The column-vectors $p_0$ and $p_1$ represent the polynomial coefficients as a single object. As we will see below, the presented considerations for one-bit logical functions allow us to obtain non-trivial results in the multi-bit case.

The combination of the basis column-vectors $e_0$ and $e_1$ form identity matrix $I = [e_0, e_1] = \begin{pmatrix} 1 & 0 \\ 0 & 1 \end{pmatrix}$. Similarly, the combination of the column-vectors $p_0$ and $p_1$ form the following matrix $P = [p_0, p_1] = \begin{pmatrix} 1 & 0 \\ 1 & 1 \end{pmatrix}$. Basis vectors $e_0$ and $e_1$ (as well as $p_0$ and $p_1$) form the basis for the representation of multi-bit functions. We will be modeling such consideration using Zhegalkin polynomials and the corresponding quantum circuits. We will see that the use of basis vectors allows us to obtain the Zhegalkin polynomial in analytic form. The use of column-vectors $p_0$ and $p_1$ makes it possible to automatize the process of finding Zhegalkin polynomials (without a lot of arithmetic actions).

## 3. ZHEGALKIN POLYNOMIALS AND QUANTUM CIRCUITS

Let us consider two-bit Boolean functions. Two-bit basis vectors are defined by tensor product of one-bit vectors. For example:

$$e_{00} = e_0 \otimes e_0 = \begin{pmatrix} 1 \\ 0 \end{pmatrix} \otimes \begin{pmatrix} 1 \\ 0 \end{pmatrix} = \begin{pmatrix} 1 \\ 0 \\ 0 \\ 0 \end{pmatrix}.$$

On the other side, $e_0 = 1 + x_1$ for the first bit, and $e_0 = 1 + x_2$ for the second bit. We immediately obtain the following Zhegalkin polynomial:

$$e_{00} = (1+x_1)(1+x_2) = 1 + x_2 + x_1 + x_1 x_2 = x_1^0 x_2^0 + x_1^0 x_2^1 + x_1^1 x_2^0 + x_1^1 x_2^1.$$

On the right side, we presented the Zhegalkin polynomial in lexicographical degree order (00, 01, 10, 11).

Using column-vector $p_0$ and tensor product $p_0 \otimes p_0$ we obtain the same result for the coefficients of the Zhegalkin polynomial: $p_{00} = p_0 \otimes p_0 = \begin{pmatrix}1\\1\end{pmatrix} \otimes \begin{pmatrix}1\\1\end{pmatrix} = \begin{pmatrix}1\\1\\1\\1\end{pmatrix}$. Note that, high dimensional tensor products are easily calculated using classical computers. Thus, vector $e_{00}$ corresponds to the column $e_{00} = \begin{pmatrix}1\\0\\0\\0\end{pmatrix}$ in the truth table, and Zhegalkin polynomial $e_{00} = 1 + x_2 + x_1 + x_1 x_2 = x_1^0 x_2^0 + x_1^0 x_2^1 + x_1^1 x_2^0 + x_1^1 x_2^1$ with coefficients $p_{00} = \begin{pmatrix}1\\1\\1\\1\end{pmatrix}$ simultaneously.

Three-qubit quantum circuit corresponds to this polynomial is shown in Fig. 3 (left). The former two qubits correspond to the function's input, the third qubit corresponds to the output. The term $x_1 x_2$ corresponds to the CCNOT gate (two first qubits control last qubit); the term $x_1$ corresponds to the CNOT gate (first qubit control last qubit); the term $x_2$ corresponds to the CNOT gate (second qubit control last qubit); finally, the term $1$ corresponds to the one-qubit X gate (NOT) acting on the last qubit.

We obtain the following equations for the last three remaining basis vectors:

$$e_{01} = e_0 \otimes e_1 = \begin{pmatrix}1\\0\end{pmatrix} \otimes \begin{pmatrix}0\\1\end{pmatrix} = \begin{pmatrix}0\\1\\0\\0\end{pmatrix} = (1+x_1)x_2 = x_2 + x_1 x_2 = 0 \cdot x_1^0 x_2^0 + 1 \cdot x_1^0 x_2^1 + 0 \cdot x_1^1 x_2^0 + 1 \cdot x_1^1 x_2^1, \quad p_{01} = p_0 \otimes p_1 = \begin{pmatrix}1\\1\end{pmatrix} \otimes \begin{pmatrix}0\\1\end{pmatrix} = \begin{pmatrix}0\\1\\0\\1\end{pmatrix}$$

$$e_{10} = e_1 \otimes e_0 = \begin{pmatrix}0\\1\end{pmatrix} \otimes \begin{pmatrix}1\\0\end{pmatrix} = \begin{pmatrix}0\\0\\1\\0\end{pmatrix} = x_1(1+x_2) = x_1 + x_1 x_2 = 0 \cdot x_1^0 x_2^0 + 0 \cdot x_1^0 x_2^1 + 1 \cdot x_1^1 x_2^0 + 1 \cdot x_1^1 x_2^1, \quad p_{10} = p_1 \otimes p_0 = \begin{pmatrix}0\\1\end{pmatrix} \otimes \begin{pmatrix}1\\1\end{pmatrix} = \begin{pmatrix}0\\0\\1\\1\end{pmatrix}$$

$$e_{11} = e_1 \otimes e_1 = \begin{pmatrix}0\\1\end{pmatrix} \otimes \begin{pmatrix}0\\1\end{pmatrix} = \begin{pmatrix}0\\0\\0\\1\end{pmatrix} = x_1 x_2 = 0 \cdot x_1^0 x_2^0 + 0 \cdot x_1^0 x_2^1 + 0 \cdot x_1^1 x_2^0 + 1 \cdot x_1^1 x_2^1, \quad p_{11} = p_1 \otimes p_1 = \begin{pmatrix}0\\1\end{pmatrix} \otimes \begin{pmatrix}0\\1\end{pmatrix} = \begin{pmatrix}0\\0\\0\\1\end{pmatrix}$$

As we can see, the column-vectors $p_{00}$, $p_{01}$, $p_{10}$ and $p_{11}$ define the coefficients of the Zhegalkin polynomial corresponding to the basis two-qubit functions $e_{00}, e_{01}, e_{10}$ and $e_{11}$ respectively. The column-vector $p_{00}$ defines the polynomial $e_{00}$, the column-vector $p_{01}$ defines the polynomial and $e_{01}$ etc. All described functions are shown in the Fig. 3.

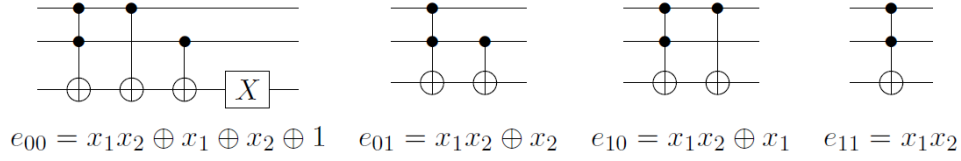

$e_{00} = x_1 x_2 \oplus x_1 \oplus x_2 \oplus 1 \quad e_{01} = x_1 x_2 \oplus x_2 \quad e_{10} = x_1 x_2 \oplus x_1 \quad e_{11} = x_1 x_2$

Figure 3. Quantum circuits for the two-qubits basis functions.

There are 16 two-bits Boolean functions (Table 2). We describe 4 basis functions. All other 12 functions can be represented as the superposition of the four basis functions:

$$e_{00} + e_{01} = \begin{pmatrix} 1 \\ 1 \\ 0 \\ 0 \end{pmatrix} = 1 + x_2 + x_1 + x_1 x_2 + x_2 + x_1 x_2 = 1 + x_1 = 1 \cdot x_1^0 x_2^0 + 0 \cdot x_1^0 x_2^1 + 1 \cdot x_1^1 x_2^0 + 0 \cdot x_1^1 x_2^1$$

We can obtain the same result in terms of $p$-vectors: $p_{00} + p_{01} = \begin{pmatrix} 1 \\ 1 \\ 1 \\ 1 \end{pmatrix} + \begin{pmatrix} 0 \\ 1 \\ 0 \\ 1 \end{pmatrix} = \begin{pmatrix} 1 \\ 0 \\ 1 \\ 0 \end{pmatrix}$. The obtained gate $e_{00} + e_{01} = 1 + x_1$ can be transformed to the CNOT gate action (first qubit control third qubit) and X gate action (last qubit).

Table 2. The truth table for 16 two-bits Boolean functions

| $x_1$ | $x_2$ | $e_{00}$ | $e_{01}$ | $e_{10}$ | $e_{11}$ | $e_{00}+e_{01}$ | $e_{00}+e_{10}$ | $e_{00}+e_{11}$ | $e_{01}+e_{10}$ |
|---|---|---|---|---|---|---|---|---|---|
| 0 | 0 | 1 | 0 | 0 | 0 | 1 | 1 | 1 | 0 |
| 0 | 1 | 0 | 1 | 0 | 0 | 1 | 0 | 0 | 1 |
| 1 | 0 | 0 | 0 | 1 | 0 | 0 | 1 | 0 | 1 |
| 1 | 1 | 0 | 0 | 0 | 1 | 0 | 0 | 1 | 0 |

| $x_1$ | $x_2$ | $e_{01}+e_{11}$ | $e_{10}+e_{11}$ | $e_{00}+1$ | $e_{01}+1$ | $e_{10}+1$ | $e_{11}+1$ | 1 | 0 |
|---|---|---|---|---|---|---|---|---|---|
| 0 | 0 | 1 | 0 | 0 | 1 | 1 | 1 | 1 | 0 |
| 0 | 1 | 0 | 0 | 1 | 0 | 1 | 1 | 1 | 0 |
| 1 | 0 | 0 | 1 | 1 | 1 | 0 | 1 | 1 | 0 |
| 1 | 1 | 1 | 1 | 1 | 1 | 1 | 0 | 1 | 0 |

The obtained results can be interpreted in terms of the following two statements.

Statement 2 (Construction algorithm of the Boolean function Zhegalkin polynomial). Arbitrary Boolean function is a superposition of the basis functions $e_{j_1 j_2 \ldots j_n} = e_{j_1} \otimes e_{j_2} \otimes \ldots \otimes e_{j_n}$. Index $j_k$ corresponds to the factor $(x_{j_k} + 1)$, where $j_k = 0$ and factor $x_{j_k}$, where $j_k = 1$. The column of the Zhegalkin polynomial coefficients is defined by sum of the $p$-columns tensor products $p_{j_1 j_2 \ldots j_n} = p_{j_1} \otimes p_{j_2} \otimes \ldots \otimes p_{j_n}$. Index $j_k$ corresponds to the column $p_0$, where $j_k = 0$ and $p_1$ where $j_k = 1$.

Statement 3 (Zhegalkin polynomials transformation to the quantum circuits). Each Zhegalkin polynomial corresponds to the certain quantum circuit. Term 1 corresponds to the $X$ gate, acting on output qubit $y$; term $x_{j_k}$ corresponds to the CNOT gate with $j_k$ control qubit; multiplication of $m$ Boolean arguments $x_{j_1} x_{j_2} ... x_{j_m}$ corresponds to the C$^{(m)}$NOT gate, where $m = 1, ..., n$, $n$ - number of input qubits.

It follows from statement 3 that a certain gate is associated with the specific term in the Zhegalkin polynomial form. We list all possible gates of this kind. One $X$ gate, acting on output qubit, $n$ different CNOT gates, $C_n^2$ C$^{(2)}$NOT gates and etc. Thus, the full set of the basis gates contains $\sum_{k=0}^{n} C_n^k = 2^n$ elements. Each of the elements can be included into the quantum circuit. Therefore, there are $2^{(2^n)}$ different Boolean quantum circuits. It corresponds to the number of possible Boolean functions.

## 4. MULTIPLE OUTPUT BOOLEAN FUNCTIONS

Before that, we have assumed that input register $x$ contains an arbitrary number $n$ of qubits, and an output register $y$ contains only one qubit. Our consideration can be easily generalized for $m$ qubits in output register. As examples we consider a half-adder and an adder. These are more complex logical elements (gates). The considered elements are usually used to implement devices with sum operation. Both elements have a two-qubit output register ($m = 2$); the half-adder has two qubits ($n = 2$) input; and the adder has three qubits ($n = 3$) input.

The states of the qubits $y_1$ and $y_2$, and the half-adder output are given by the following equations:

$$y_1 = x_1 + x_2, \; y_2 = x_1 x_2.$$

The similar equations are given for the adder element:

$$y_1 = x_1 + x_2 + x_3, \; y_2 = x_1 x_2 + x_1 x_3 + x_2 x_3.$$

There are the quantum circuits of the half-adder and adder below.

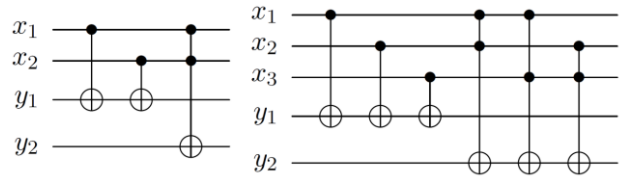

Figure 4. Quantum circuits for the half-adder (left) and adder (right)..

The statement 1 can be generalized to the multi-bit output case. As described earlier for one-bit input register, there are two $2 \times 2$ matrices $u_0 = I$ and $u_1 = X$. In the general case ($m$-bit output register), there are $2^m \times 2^m$ matrices. For example, for two-bit output register we obtain four $4 \times 4$ matrices: $u_{00} = u_0 \otimes u_0 = I \otimes I$, $u_{01} = u_0 \otimes u_1 = I \otimes X$, $u_{10} = u_1 \otimes u_0 = X \otimes I$, $u_{11} = u_1 \otimes u_1 = X \otimes X$. Here, the bottom indices define the bits values of the Boolean function output. It makes the block matrix construction obvious in more complex cases. Truth tables for the half-adder and adder are presented in Table 3 and Table 4.

Table 3. Half-adder truth table.

| $x_1$ | $x_2$ | $y_1 = x_1 + x_2$ | $y_2 = x_1 x_2$ |
|---|---|---|---|
| 0 | 0 | 0 | 0 |
| 0 | 1 | 1 | 0 |
| 1 | 0 | 1 | 0 |
| 1 | 1 | 0 | 1 |

Table 4. Adder truth table.

| $x_1$ | $x_2$ | $x_3$ | $y_1 = x_1 + x_2 + x_3$ | $y_2 = x_1x_2 + x_1x_3 + x_2x_3$ |
|---|---|---|---|---|
| 0 | 0 | 0 | 0 | 0 |
| 0 | 0 | 1 | 1 | 0 |
| 0 | 1 | 0 | 1 | 0 |
| 0 | 1 | 1 | 0 | 1 |
| 1 | 0 | 0 | 1 | 0 |
| 1 | 0 | 1 | 0 | 1 |
| 1 | 1 | 0 | 0 | 1 |
| 1 | 1 | 1 | 1 | 1 |

The presented methods with truth tables allow us to construct the unitary transform matrices for arbitrary Boolean functions with $n$-bit input and $m$-bit output. It is not difficult to construct unitary transform matrices $U_f$ for the half-adder and adder. Note that, the half-adder matrix has dimension $16 \times 16$, and the adder matrix has dimension $32 \times 32$. In the general case, the Boolean function with $n$-bit input and $m$-bit output corresponds to the unitary matrix $U_f \left( 2^{n+m} \times 2^{n+m} \right)$ acting on the $(n+m)$-qubit register. Therefore, we can generalize the statement 1.

<u>Statement 1A</u> (on the block-diagonal character of Boolean transforms with $n$-bit input and $m$-bit output). The unitary matrix corresponding to the Boolean function has a block-diagonal form, and the function values $f(x) = j_1 ... j_m$ correspond to the matrix $u_{j_1...j_m} = u_{j_1} \otimes ... \otimes u_{j_m}$, where $u_0 = I$, and $u_1 = X$.

Note that, the $m$ Boolean columns (values of the Boolean function with $m$-qubit output) appear in the truth table. The statements 2 and 3 are directly applicable to these columns. This property has been used by us to construct the quantum circuits of the half-adder and adder.

## 5. THREE-VALUED LOGIC

Three-leveled systems (qutrits) are used in three-valued logic (instead of traditional two-leveled qubits). There are $3^{(3^n)}$ various functions, where $n$ – the number of basis three-leveled elements (qutrits) in output register.

$n = 1$: $3^3 = 27$ Boolean functions;

$n = 2$: $3^{(3^2)} = 3^9 = 19683$ Boolean functions;

$n = 3$: $3^{(3^3)} = 3^{27} = 7.625.597.484.987 \approx 7.6 \cdot 10^{12}$ Boolean functions.

The unitary transforms matrices corresponding to the logical shifts (0,1 and 2 respectively) are the following:

$$T_0 = I = \begin{pmatrix} 1 & 0 & 0 \\ 0 & 1 & 0 \\ 0 & 0 & 1 \end{pmatrix}, \quad T_1 = \begin{pmatrix} 0 & 0 & 1 \\ 1 & 0 & 0 \\ 0 & 1 & 0 \end{pmatrix}, \quad T_2 = \begin{pmatrix} 0 & 1 & 0 \\ 0 & 0 & 1 \\ 1 & 0 & 0 \end{pmatrix}.$$

There are the necessary requirements for the $T_1$ and $T_2$ transforms. We will consider all operations mod 3. The $T_1$ matrix provides a shift ($|0\rangle \to |1\rangle, |1\rangle \to |2\rangle, |2\rangle \to |0\rangle$). We can represent it using Dirac notations $T_1 = |1\rangle\langle 0| + |2\rangle\langle 1| + |0\rangle\langle 2|$. Similarly, the $T_2$ matrix provides a shift ($|0\rangle \to |2\rangle, |1\rangle \to |0\rangle, |2\rangle \to |1\rangle$). Here, we have $T_2 = |2\rangle\langle 0| + |0\rangle\langle 1| + |1\rangle\langle 2|$. It is easy to verify the validity of the following equations: $T_1^2 = T_2$, $T_1^3 = T_0 = I$.

The presented results are consistent with the idea of the defined operators. The $T_2$ operator provides the shift by 2. It corresponds to the double action of the $T_1$ operator providing the shift by 1. Threefold application of the $T_1$ operator leads to shift by 3 (it is equivalent to the identity transformation). As can be seen, the operators action is similar to a 120 degree plane rotation. Three such sequentially rotations lead to the identity transform. It turns out that such visual geometric interpretation can be strictly formalized. Let $T = \exp(-iS\theta)$. $\theta$ – rotation angle, $S$ - Hermitian operator. We need to select $S$ operator with the property: $S = T_1$, when $\theta = \frac{2\pi}{3}$ (120 degree). Therefore, $\log(T_1) = -i\frac{2\pi}{3}S$. It follows to the equation:

$$S = \frac{i}{\sqrt{3}}\begin{pmatrix} 0 & -1 & 1 \\ 1 & 0 & -1 \\ -1 & 1 & 0 \end{pmatrix}.$$

The resulting $S$ matrix is the Hermitian operator. This operator defines the spin 1 projection to some particular direction. It can be shown that 120 degree rotation ($\theta = \frac{2\pi}{3}$) leads to $T_1$ operator. Similarly, the 240 degree rotation leads to $T_2$ operator. Zero and 360 degree angles lead to the identity transform. $S$ operator eigenvalues are $m = -1, 0, 1$. These numbers correspond to the possible spin projections. Thus, they enumerate the possible states corresponding to the spin 1 particles (three-leveled logical element in our case). The traditional number of the logical state is different from the spin projection per unit: $x = m + 1 = 0, 1, 2$. Note that the presented matrix exponent is useful for an arbitrary rotation angle $\theta$. In the general case, the system transformed to all three basis states superposition. Let, the initial system state be $|0\rangle$. Let us consider a 60 degree ($\theta = \frac{\pi}{3}$) angle transform. This transform can be called the "1/2 shift" because it corresponds to the half of the shift 1 angle. It can be shown that such transform leads to superposition state $|\psi\rangle = \frac{2}{3}|0\rangle + \frac{2}{3}|1\rangle - \frac{1}{3}|2\rangle$. The obtained coefficients are defined as the probability amplitudes. The considered general transform corresponds to the transition from traditional discrete mathematics to quantum computer science.

The truth table of a given function allows us to easily construct its unitary representation (3x3 block-diagonal form).

<u>Statement 1B</u> (on the block-diagonal character of Boolean transforms for three-valued logic). The unitary matrix corresponding to the Boolean function has a block-diagonal form. Function values $f(x) = 0$ correspond to the $T_0 = I$ matrix; values $f(x) = 1$ correspond to the $T_1$ matrix; and values $f(x) = 2$ correspond to the $T_2$ matrix.

It is necessary to consider all degrees (from 0 to 2) for the basis functions obtaining (we consider all arithmetic operations mod 3).

$$x = \begin{pmatrix} 0 \\ 1 \\ 2 \end{pmatrix}, \quad x^0 = \begin{pmatrix} 1 \\ 1 \\ 1 \end{pmatrix}, \quad x^1 = x = \begin{pmatrix} 0 \\ 1 \\ 2 \end{pmatrix}, \quad x^2 = \begin{pmatrix} 0 \\ 1 \\ 1 \end{pmatrix}.$$

Let us combine column-vectors $x^0$, $x^1$ and $x^2$ into a single matrix: $Q = \begin{bmatrix} x^0 & x^1 & x^2 \end{bmatrix} = \begin{pmatrix} 1 & 0 & 0 \\ 1 & 1 & 1 \\ 1 & 2 & 1 \end{pmatrix}$.

Standard basis functions can be represented as a columns $x^0$, $x^1$ and $x^2$ superposition:

$$e_0 = \begin{pmatrix} 1 \\ 0 \\ 0 \end{pmatrix} = 1 + 2x^2 = 1 \cdot x^0 + 0 \cdot x^1 + 2x^2, \quad e_1 = \begin{pmatrix} 0 \\ 1 \\ 0 \end{pmatrix} = 2x + 2x^2 = 0 \cdot x^0 + 2 \cdot x^1 + 2x^2, \quad e_2 = \begin{pmatrix} 0 \\ 0 \\ 1 \end{pmatrix} = x + 2x^2 = 0 \cdot x^0 + 1 \cdot x^1 + 2x^2.$$

In full analogy with two-valued logic discussed above, we define column-vectors $p_0$, $p_1$ and $p_2$, containing the Zhegalkin polynomials coefficients for the functions $e_0, e_1$ and $e_2$ respectively.

$$p_0 = \begin{pmatrix} 1 \\ 0 \\ 2 \end{pmatrix}, \quad p_1 = \begin{pmatrix} 0 \\ 2 \\ 2 \end{pmatrix}, \quad p_2 = \begin{pmatrix} 0 \\ 1 \\ 2 \end{pmatrix}.$$

Let us combine all considered column-vectors into the matrix: $P = \begin{bmatrix} p_0 & p_1 & p_2 \end{bmatrix} = \begin{pmatrix} 1 & 0 & 0 \\ 0 & 2 & 1 \\ 2 & 2 & 2 \end{pmatrix}$.

The considered matrix $P$ is inverse to the matrix $Q$: $QP = PQ = I$. Note that the matrix has the following form (for the two-leveled logic): $Q = \begin{bmatrix} x^0 & x^1 \end{bmatrix} = \begin{pmatrix} 1 & 0 \\ 1 & 1 \end{pmatrix}$. For the two-leveled logic: $Q = P$. At the same time, the identity $QP = PQ = I$ is satisfied. This identity turns into an identity $Q^2 = I$ (for the considered case).

The detailed calculation example of the three-valued logic Zhegalkin polynomial is presented in Appendix 1.

## 6. THE GENERAL ALGORITHM FOR THE CONSTRUCTION OF A BOOLEAN FUNCTION IN THE ZHEGALKIN POLYNOMIAL FORM

In the general case ($k$-valued logic) it is necessary to consider all degrees of $x$ from 0 to $k-1$ to construct basis functions. The corresponding set will be complete if and only if $k = p$, $p$ - prime number [2]. Otherwise, the consideration is equivalent to the k=3 case.

We consider the functions with $n$ arguments for the $k$-leveled logic, $k$ is a prime number. Let the logical function be a column with $k^n$ numbers in lexicographical order. Each of the $k^n$ logical value is an integer from 0 to $k-1$. Let the logic values column be $x = \begin{pmatrix} 0 \\ 1 \\ \vdots \\ k-1 \end{pmatrix}$. We can construct a $k \times k$ matrix $Q = \begin{bmatrix} x^0 & \ldots & x^{k-1} \end{bmatrix}$ based on $x$ degrees from 0 to k – 1. Inverse matrix $P$ is constructed from the $Q$ matrix. Thus, the equation $QP = PQ = I$ holds true. To calculate the Zhegalkin polynomial coefficients we need to use columns of the $P$ matrix: $P = \begin{bmatrix} p_0 & ... & p_{k-1} \end{bmatrix}$. Note that $k$ is a prime number. Such condition defines the existence of inverse matrix for the $Q$ matrix.

<u>Statement 2 in general case</u> (algorithm for the construction a Boolean function in a Zhegalkin polynomial form). The column $a$ with dimension $k^n$ defines the Zhegalkin polynomial coefficients obtained by $p$-columns $p_{j_1 j_2 ... j_n} = p_{j_1} \otimes p_{j_2} \otimes ... \otimes p_{j_n}$ tensor products sum over all rows in the truth table with weights equal to logic function $f$ values. Logical index $j_m$ ($j_m = 0, 1, .., k-1$) corresponds to the column-vector $p_{j_m}$. In other words, we obtain $p_0$ where $j_m = 0$ and $p_1$ where $j_m = 1$, ..., $p_{k-1}$ where $j_m = k-1$.

The resulting column-vector $a$ defines the Zhegalkin polynomial coefficients. This vector is represented into degrees $x_1, ..., x_n$ lexicographical order.

The presented statement defines the transformation algorithm from logic function $f$ column into a column-vector $a$ with Zhegalkin polynomial coefficient. Here, the $P$ matrix columns are used with weights corresponding to logical function $f$ values.

It can be shown that the inverse transform (from column-vector $a$ to logical function $f$ values) corresponds to the $Q$ matrix columns using (with $a$ vector elements as weights).

The sequential application of the considered transforms (direct and inverse) leads to the identity transform. Thus, there is a dualism between the column-functions $f$ and $a$: $f \xrightleftharpoons[Q]{P} a$, or $a = P * f$, $f = Q * a$. The first equation defines $P$-transform, the second equation defines $Q$-transform. $P$-transform allows us to find the column-vector with the Zhegalkin polynomial coefficients $a$ from known Boolean function $f$. $Q$-transform allows us to find the Boolean function $f$ from known column-vector with Zhegalkin polynomial coefficients $a$.

The presented methods and algorithms have been tested for different $k$-logics (five-level logic example is presented in Appendix 1), $k = 2, 3, 5, 7, 11, 13, 17, 19, 23$ and $29$. Correspondings $P$ matricies are presented in Appendix 2.

## 7. CONCLUSIONS

The presented methods and algorithms demonstrate the deep relationship between the classical discrete mathematics and quantum information science. Developed methods and algorithms allow us to consider the unitary quantum representations for an arbitrary Boolean functions with multi-bit input and multi-bit output. The relationship between Zhegalkin polynomials and quantum logic circuits is described in details. The developed theory is generalized to the multi-level ($k$-level) logics, where $k = p$, $p$ - prime number.

The developed methods and algorithms can significantly improve the transition from classical computer logic to quantum computers.

## ACKNOWLEDGMENTS

The work is supported by Russian Science Foundation (RSF), project no: 14-12-01338П.

# APPENDIX 1. ZHEGALKIN POLYNOMIAL CALCULATION IN MULTI-LEVEL LOGICS

**Three-valued logic example**

Let consider the following example: the three-valued Boolean function with two arguments defined by truth table.

| $x_1$ | $x_2$ | $f$ |
|---|---|---|
| 0 | 0 | 0 |
| 0 | 1 | 0 |
| 0 | 2 | 2 |
| 1 | 0 | 0 |
| 1 | 1 | 1 |
| 1 | 2 | 0 |
| 2 | 0 | 2 |
| 2 | 1 | 0 |
| 2 | 2 | 0 |

You can easily calculate the Zhegalkin polynomial for this function. Only non-zero rows of the truth table should be considered, therefore:

$$f = 2e_0 e_2 + e_1 e_1 + 2e_2 e_0 = 2(2x_1^2 + 1)(2x_2^2 + x_2) + (2x_1^2 + 2x_1)(2x_2^2 + 2x_2) + 2(2x_1^2 + x_1)(2x_2^2 + 1) =$$
$$= 2x_2 + x_2^2 + 2x_1 + x_1 x_2 + 2x_1 x_2^2 + x_1^2 + 2x_1^2 x_2 + 2x_1^2 x_2^2$$

Our numerical calculation confirms the correctness of the expression.

Using $p$-columns terminology, we obtain the same result for the column of the Zhegalkin polynomial coefficients $a$ using tensor products apparatus. Such approach is much simpler and can be easily implemented numerically in the case of high dimensions.

$$a = 2p_0 p_2 + p_1 p_1 + 2p_2 p_0 = 2\begin{pmatrix}1\\0\\2\end{pmatrix} \otimes \begin{pmatrix}0\\1\\2\end{pmatrix} + \begin{pmatrix}0\\2\\2\end{pmatrix} \otimes \begin{pmatrix}0\\2\\2\end{pmatrix} + 2\begin{pmatrix}0\\1\\2\end{pmatrix} \otimes \begin{pmatrix}1\\0\\2\end{pmatrix} = \begin{pmatrix}0\\2\\1\\2\\1\\2\\1\\2\\2\end{pmatrix}$$

Using the $Q$ matrix obtained in section 5 for three-valued logic, we can apply the reverse transform: from Zhegalkin polynomial coefficients to the Boolean function vector. In this way

$$f = (Q \otimes Q) * a.$$

Such transforms based on $P$ and $Q$ matrices allows us move from Boolean function vector to the Zhegalkin polynomials coefficients and vice versa.

**Five-valued logic example**

We need to consider degrees from $0$ to $k-1$ for the basis functions construction if we use $k$–level logic. The corresponding set will be complete if and only if $k = p$ where $p$ is a prime number [2]. In other cases, the consideration approach is similar to the $k = 3$ case. Let consider five-valued logic. In this case, we need to consider the $x$ degrees from $0$ to $4$.

$$x = \begin{pmatrix} 0 \\ 1 \\ 2 \\ 3 \\ 4 \end{pmatrix}, \quad x^0 = \begin{pmatrix} 1 \\ 1 \\ 1 \\ 1 \\ 1 \end{pmatrix}, \quad x^1 = x = \begin{pmatrix} 0 \\ 1 \\ 2 \\ 3 \\ 4 \end{pmatrix}, \quad x^2 = \begin{pmatrix} 0 \\ 1 \\ 4 \\ 4 \\ 1 \end{pmatrix}, \quad x^3 = \begin{pmatrix} 0 \\ 1 \\ 3 \\ 2 \\ 4 \end{pmatrix}, \quad x^4 = \begin{pmatrix} 0 \\ 1 \\ 1 \\ 1 \\ 1 \end{pmatrix}$$

Then, the standard basic functions can be represented as:

$$e_0 = \begin{pmatrix} 1 \\ 0 \\ 0 \\ 0 \\ 0 \end{pmatrix} = 4x^4 + 1, \quad e_1 = \begin{pmatrix} 0 \\ 1 \\ 0 \\ 0 \\ 0 \end{pmatrix} = 4x^4 + 4x^3 + 4x^2 + 4x, \quad e_2 = \begin{pmatrix} 0 \\ 0 \\ 1 \\ 0 \\ 0 \end{pmatrix} = 4x^4 + 3x^3 + x^2 + 2x,$$

$$e_3 = \begin{pmatrix} 0 \\ 0 \\ 0 \\ 1 \\ 0 \end{pmatrix} = 4x^4 + 2x^3 + x^2 + 3x, \quad e_4 = \begin{pmatrix} 0 \\ 0 \\ 0 \\ 0 \\ 1 \end{pmatrix} = 4x^4 + x^3 + 4x^2 + x$$

The corresponding matrix $p$ to the considered vectors is:

$$P = \begin{pmatrix} 1 & 0 & 0 & 0 & 0 \\ 0 & 4 & 2 & 3 & 1 \\ 0 & 4 & 1 & 1 & 4 \\ 0 & 4 & 3 & 2 & 1 \\ 4 & 4 & 4 & 4 & 4 \end{pmatrix}$$

We can easily obtain the $a$ vector with Zhegalkin polynomial coefficients using this matrix for arbitrary five-valued Boolean function. Let's take for an example the following function $f = (0, 2, 1, 0, 2)$. Therefore, the Zhegalkin polynomial is

$$a = P * f = \begin{pmatrix} 1 & 0 & 0 & 0 & 0 \\ 0 & 4 & 2 & 3 & 1 \\ 0 & 4 & 1 & 1 & 4 \\ 0 & 4 & 3 & 2 & 1 \\ 4 & 4 & 4 & 4 & 4 \end{pmatrix} * \begin{pmatrix} 0 \\ 2 \\ 1 \\ 0 \\ 2 \end{pmatrix} = \begin{pmatrix} 0 \\ 2 \\ 2 \\ 3 \\ 0 \end{pmatrix}$$

Thus, this method allows us to calculate the Zhegalkin polynomial coefficients in two different ways. Both these methods significantly speed up the calculation process of high-dimensional vector describes the Zhegalkin coefficients.

Let us consider the five-logic Boolean function of two arguments presented in the truth table. Let $f(0,3)=2$, $f(2,1)=4$. Considered function turn to zero at the all remaining points.

| $x_1$ | 0 | 0 | 0 | 0 | 0 | 1 | 1 | 1 | 1 | 1 | 2 | 2 | 2 | 2 | 2 | 3 | 3 | 3 | 3 | 3 | 4 | 4 | 4 | 4 | 4 |
|---|---|---|---|---|---|---|---|---|---|---|---|---|---|---|---|---|---|---|---|---|---|---|---|---|---|
| $x_2$ | 0 | 1 | 2 | 3 | 4 | 0 | 1 | 2 | 3 | 4 | 0 | 1 | 2 | 3 | 4 | 0 | 1 | 2 | 3 | 4 | 0 | 1 | 2 | 3 | 4 |
| $f$ | 0 | 0 | 0 | 2 | 0 | 0 | 0 | 0 | 0 | 0 | 0 | 4 | 0 | 0 | 0 | 0 | 0 | 0 | 0 | 0 | 0 | 0 | 0 | 0 | 0 |

The Zhegalkin polynomial calculation for the considered function is presented below. As is an three-valued logic example, we need to consider only non-zero rows of the truth table.

$$f = 2e_0e_3 + 4e_2e_1 = x_1^4 x_2^4 + 2x_1^4 x_2^2 + 3x_1^4 x_2 + 3x_1^3 x_2^4 + 3x_1^3 x_2^3 + 3x_1^3 x_2^2 + 3x_1^3 x_2 + x_1^2 x_2^4 +$$
$$+ x_1^2 x_2^3 + x_1^2 x_2^2 + 2x_1 x_2^4 + 2x_1 x_2^3 + 2x_1 x_2^2 + x_1^2 x_2 + 2x_1 x_2 + 3x_2^4 + 4x_2^3 + 2x_2^2 + x_2;$$

Our numerical calculation confirms the correctness of the expression. It is easy to obtain the same result using $p$-columns terminology or $P$ matrix terminology

$$a = (P \otimes P) * f.$$

Therefore, $a = (0,1,2,4,3,0,2,2,2,2,0,1,1,1,1,0,3,3,3,3,0,3,2,0,1)$ for considered example.

## APPENDIX 2. *P* MATRICIES FOR DIFFERENT *k* VALUES

$k = 2$:

$$P = \begin{pmatrix} 1 & 0 \\ 1 & 1 \end{pmatrix}$$

$k = 3$:

$$P = \begin{pmatrix} 1 & 0 & 0 \\ 0 & 2 & 1 \\ 2 & 2 & 2 \end{pmatrix}$$

$k = 5$:

$$P = \begin{pmatrix} 1 & 0 & 0 & 0 & 0 \\ 0 & 4 & 2 & 3 & 1 \\ 0 & 4 & 1 & 1 & 4 \\ 0 & 4 & 3 & 2 & 1 \\ 4 & 4 & 4 & 4 & 4 \end{pmatrix}$$

$k = 7$:

$$P = \begin{pmatrix} 1 & 0 & 0 & 0 & 0 & 0 & 0 \\ 0 & 6 & 3 & 2 & 5 & 4 & 1 \\ 0 & 6 & 5 & 3 & 3 & 5 & 6 \\ 0 & 6 & 6 & 1 & 6 & 1 & 1 \\ 0 & 6 & 3 & 5 & 5 & 3 & 6 \\ 0 & 6 & 5 & 4 & 3 & 2 & 1 \\ 6 & 6 & 6 & 6 & 6 & 6 & 6 \end{pmatrix}$$

$k = 11$:

$$P = \begin{pmatrix} 1 & 0 & 0 & 0 & 0 & 0 & 0 & 0 & 0 & 0 & 0 \\ 0 & 10 & 5 & 7 & 8 & 2 & 9 & 3 & 4 & 6 & 1 \\ 0 & 10 & 8 & 6 & 2 & 7 & 7 & 2 & 6 & 8 & 10 \\ 0 & 10 & 4 & 2 & 6 & 8 & 3 & 5 & 9 & 7 & 1 \\ 0 & 10 & 2 & 8 & 7 & 6 & 6 & 7 & 8 & 2 & 10 \\ 0 & 10 & 1 & 10 & 10 & 10 & 1 & 1 & 1 & 10 & 1 \\ 0 & 10 & 6 & 7 & 8 & 2 & 2 & 8 & 7 & 6 & 10 \\ 0 & 10 & 3 & 6 & 2 & 7 & 4 & 9 & 5 & 8 & 1 \\ 0 & 10 & 7 & 2 & 6 & 8 & 8 & 6 & 2 & 7 & 10 \\ 0 & 10 & 9 & 8 & 7 & 6 & 5 & 4 & 3 & 2 & 1 \\ 10 & 10 & 10 & 10 & 10 & 10 & 10 & 10 & 10 & 10 & 10 \end{pmatrix}$$

$k = 13$:

$$P = \begin{pmatrix} 1 & 0 & 0 & 0 & 0 & 0 & 0 & 0 & 0 & 0 & 0 & 0 & 0 \\ 0 & 12 & 6 & 4 & 3 & 5 & 2 & 11 & 8 & 10 & 9 & 7 & 1 \\ 0 & 12 & 3 & 10 & 4 & 1 & 9 & 9 & 1 & 4 & 10 & 3 & 12 \\ 0 & 12 & 8 & 12 & 1 & 8 & 8 & 5 & 5 & 12 & 1 & 5 & 1 \\ 0 & 12 & 4 & 4 & 10 & 12 & 10 & 10 & 12 & 10 & 4 & 4 & 12 \\ 0 & 12 & 2 & 10 & 9 & 5 & 6 & 7 & 8 & 4 & 3 & 11 & 1 \\ 0 & 12 & 1 & 12 & 12 & 1 & 1 & 1 & 1 & 12 & 12 & 1 & 12 \\ 0 & 12 & 7 & 4 & 3 & 8 & 11 & 2 & 5 & 10 & 9 & 6 & 1 \\ 0 & 12 & 10 & 10 & 4 & 12 & 4 & 4 & 12 & 4 & 10 & 10 & 12 \\ 0 & 12 & 5 & 12 & 1 & 5 & 5 & 8 & 8 & 12 & 1 & 8 & 1 \\ 0 & 12 & 9 & 4 & 10 & 1 & 3 & 3 & 1 & 10 & 4 & 9 & 12 \\ 0 & 12 & 11 & 10 & 9 & 8 & 7 & 6 & 5 & 4 & 3 & 2 & 1 \\ 12 & 12 & 12 & 12 & 12 & 12 & 12 & 12 & 12 & 12 & 12 & 12 & 12 \end{pmatrix}$$

$k = 17$:

$$P = \begin{pmatrix}
1 & 0 & 0 & 0 & 0 & 0 & 0 & 0 & 0 & 0 & 0 & 0 & 0 & 0 & 0 & 0 & 0 \\
0 & 16 & 8 & 11 & 4 & 10 & 14 & 12 & 2 & 15 & 5 & 3 & 7 & 13 & 6 & 9 & 1 \\
0 & 16 & 4 & 15 & 1 & 2 & 8 & 9 & 13 & 13 & 9 & 8 & 2 & 1 & 15 & 4 & 16 \\
0 & 16 & 2 & 5 & 13 & 14 & 7 & 11 & 8 & 9 & 6 & 10 & 3 & 4 & 12 & 15 & 1 \\
0 & 16 & 1 & 13 & 16 & 13 & 4 & 4 & 1 & 1 & 4 & 4 & 13 & 16 & 13 & 1 & 16 \\
0 & 16 & 9 & 10 & 4 & 6 & 12 & 3 & 15 & 2 & 14 & 5 & 11 & 13 & 7 & 8 & 1 \\
0 & 16 & 13 & 9 & 1 & 8 & 2 & 15 & 4 & 4 & 15 & 2 & 8 & 1 & 9 & 13 & 16 \\
0 & 16 & 15 & 3 & 13 & 5 & 6 & 7 & 9 & 8 & 10 & 11 & 12 & 4 & 14 & 2 & 1 \\
0 & 16 & 16 & 1 & 16 & 1 & 1 & 1 & 16 & 16 & 1 & 1 & 1 & 16 & 1 & 16 & 16 \\
0 & 16 & 8 & 6 & 4 & 7 & 3 & 5 & 2 & 15 & 12 & 14 & 10 & 13 & 11 & 9 & 1 \\
0 & 16 & 4 & 2 & 1 & 15 & 9 & 8 & 13 & 13 & 8 & 9 & 15 & 1 & 2 & 4 & 16 \\
0 & 16 & 2 & 12 & 13 & 3 & 10 & 6 & 8 & 9 & 11 & 7 & 14 & 4 & 5 & 15 & 1 \\
0 & 16 & 1 & 4 & 16 & 4 & 13 & 13 & 1 & 1 & 13 & 13 & 4 & 16 & 4 & 1 & 16 \\
0 & 16 & 9 & 7 & 4 & 11 & 5 & 14 & 15 & 2 & 3 & 12 & 6 & 13 & 10 & 8 & 1 \\
0 & 16 & 13 & 8 & 1 & 9 & 15 & 2 & 4 & 4 & 2 & 15 & 9 & 1 & 8 & 13 & 16 \\
0 & 16 & 15 & 14 & 13 & 12 & 11 & 10 & 9 & 8 & 7 & 6 & 5 & 4 & 3 & 2 & 1 \\
16 & 16 & 16 & 16 & 16 & 16 & 16 & 16 & 16 & 16 & 16 & 16 & 16 & 16 & 16 & 16 & 16
\end{pmatrix}$$

$k = 19$:

$$P = \begin{pmatrix}
1 & 0 & 0 & 0 & 0 & 0 & 0 & 0 & 0 & 0 & 0 & 0 & 0 & 0 & 0 & 0 & 0 & 0 & 0 \\
0 & 18 & 9 & 6 & 14 & 15 & 3 & 8 & 7 & 2 & 17 & 12 & 11 & 16 & 4 & 5 & 13 & 10 & 1 \\
0 & 18 & 14 & 2 & 13 & 3 & 10 & 12 & 8 & 15 & 15 & 8 & 12 & 10 & 3 & 13 & 2 & 14 & 18 \\
0 & 18 & 7 & 7 & 8 & 12 & 8 & 18 & 1 & 8 & 11 & 18 & 1 & 11 & 7 & 11 & 12 & 12 & 1 \\
0 & 18 & 13 & 15 & 2 & 10 & 14 & 8 & 12 & 3 & 3 & 12 & 8 & 14 & 10 & 2 & 15 & 13 & 18 \\
0 & 18 & 16 & 5 & 10 & 2 & 15 & 12 & 11 & 13 & 6 & 8 & 7 & 4 & 17 & 9 & 14 & 3 & 1 \\
0 & 18 & 8 & 8 & 12 & 8 & 12 & 18 & 18 & 12 & 12 & 18 & 18 & 12 & 8 & 12 & 8 & 8 & 18 \\
0 & 18 & 4 & 9 & 3 & 13 & 2 & 8 & 7 & 14 & 5 & 12 & 11 & 17 & 6 & 16 & 10 & 15 & 1 \\
0 & 18 & 2 & 3 & 15 & 14 & 13 & 12 & 8 & 10 & 10 & 8 & 12 & 13 & 14 & 15 & 3 & 2 & 18 \\
0 & 18 & 1 & 1 & 18 & 18 & 18 & 18 & 1 & 18 & 1 & 18 & 1 & 1 & 1 & 1 & 18 & 18 & 1 \\
0 & 18 & 10 & 13 & 14 & 15 & 3 & 8 & 12 & 2 & 2 & 12 & 8 & 3 & 15 & 14 & 13 & 10 & 18 \\
0 & 18 & 5 & 17 & 13 & 3 & 10 & 12 & 11 & 15 & 4 & 8 & 7 & 9 & 16 & 6 & 2 & 14 & 1 \\
0 & 18 & 12 & 12 & 8 & 12 & 8 & 18 & 18 & 8 & 8 & 18 & 18 & 8 & 12 & 8 & 12 & 12 & 18 \\
0 & 18 & 6 & 4 & 2 & 10 & 14 & 8 & 7 & 3 & 16 & 12 & 11 & 5 & 9 & 17 & 15 & 13 & 1 \\
0 & 18 & 3 & 14 & 10 & 2 & 15 & 12 & 8 & 13 & 13 & 8 & 12 & 15 & 2 & 10 & 14 & 3 & 18 \\
0 & 18 & 11 & 11 & 12 & 8 & 12 & 18 & 1 & 12 & 7 & 18 & 1 & 7 & 11 & 7 & 8 & 8 & 1 \\
0 & 18 & 15 & 10 & 3 & 13 & 2 & 8 & 12 & 14 & 14 & 12 & 8 & 2 & 13 & 3 & 10 & 15 & 18 \\
0 & 18 & 17 & 16 & 15 & 14 & 13 & 12 & 11 & 10 & 9 & 8 & 7 & 6 & 5 & 4 & 3 & 2 & 1 \\
18 & 18 & 18 & 18 & 18 & 18 & 18 & 18 & 18 & 18 & 18 & 18 & 18 & 18 & 18 & 18 & 18 & 18 & 18
\end{pmatrix}$$

$k = 23$:

$$P = \begin{pmatrix}
1 & 0 & 0 & 0 & 0 & 0 & 0 & 0 & 0 & 0 & 0 & 0 & 0 & 0 & 0 & 0 & 0 & 0 & 0 & 0 & 0 & 0 & 0 \\
0 & 22 & 11 & 15 & 17 & 9 & 19 & 13 & 20 & 5 & 16 & 2 & 21 & 7 & 18 & 3 & 10 & 4 & 14 & 6 & 8 & 12 & 1 \\
0 & 22 & 17 & 5 & 10 & 11 & 7 & 15 & 14 & 21 & 20 & 19 & 19 & 20 & 21 & 14 & 15 & 7 & 11 & 10 & 5 & 17 & 22 \\
0 & 22 & 20 & 17 & 14 & 16 & 5 & 12 & 19 & 10 & 2 & 8 & 15 & 21 & 13 & 4 & 11 & 18 & 7 & 9 & 6 & 3 & 1 \\
0 & 22 & 10 & 21 & 15 & 17 & 20 & 5 & 11 & 19 & 14 & 7 & 7 & 14 & 19 & 11 & 5 & 20 & 17 & 15 & 21 & 10 & 22 \\
0 & 22 & 5 & 7 & 21 & 8 & 11 & 4 & 10 & 20 & 6 & 9 & 14 & 17 & 3 & 13 & 19 & 12 & 15 & 2 & 16 & 18 & 1 \\
0 & 22 & 14 & 10 & 11 & 20 & 21 & 17 & 7 & 15 & 19 & 5 & 5 & 19 & 15 & 7 & 17 & 21 & 20 & 11 & 10 & 14 & 22 \\
0 & 22 & 7 & 11 & 20 & 4 & 15 & 9 & 21 & 17 & 18 & 13 & 10 & 5 & 6 & 2 & 14 & 8 & 19 & 3 & 12 & 16 & 1 \\
0 & 22 & 15 & 19 & 5 & 10 & 14 & 21 & 17 & 7 & 11 & 20 & 20 & 11 & 7 & 17 & 21 & 14 & 10 & 5 & 19 & 15 & 22 \\
0 & 22 & 19 & 14 & 7 & 2 & 10 & 3 & 5 & 11 & 8 & 6 & 17 & 15 & 12 & 18 & 20 & 13 & 21 & 16 & 9 & 4 & 1 \\
0 & 22 & 21 & 20 & 19 & 5 & 17 & 7 & 15 & 14 & 10 & 11 & 11 & 10 & 14 & 15 & 7 & 17 & 5 & 19 & 20 & 21 & 22 \\
0 & 22 & 22 & 22 & 22 & 1 & 22 & 1 & 22 & 22 & 1 & 1 & 22 & 22 & 1 & 1 & 22 & 1 & 22 & 1 & 1 & 1 & 1 \\
0 & 22 & 11 & 15 & 17 & 14 & 19 & 10 & 20 & 5 & 7 & 21 & 21 & 7 & 5 & 20 & 10 & 19 & 14 & 17 & 15 & 11 & 22 \\
0 & 22 & 17 & 5 & 10 & 12 & 7 & 8 & 14 & 21 & 3 & 4 & 19 & 20 & 2 & 9 & 15 & 16 & 11 & 13 & 18 & 6 & 1 \\
0 & 22 & 20 & 17 & 14 & 7 & 5 & 11 & 19 & 10 & 21 & 15 & 15 & 21 & 10 & 19 & 11 & 5 & 7 & 14 & 17 & 20 & 22 \\
0 & 22 & 10 & 21 & 15 & 6 & 20 & 18 & 11 & 19 & 9 & 16 & 7 & 14 & 4 & 12 & 5 & 3 & 17 & 8 & 2 & 13 & 1 \\
0 & 22 & 5 & 7 & 21 & 15 & 11 & 19 & 10 & 20 & 17 & 14 & 14 & 17 & 20 & 10 & 19 & 11 & 15 & 21 & 7 & 5 & 22 \\
0 & 22 & 14 & 10 & 11 & 3 & 21 & 6 & 7 & 15 & 4 & 18 & 5 & 19 & 8 & 16 & 17 & 2 & 20 & 12 & 13 & 9 & 1 \\
0 & 22 & 7 & 11 & 20 & 19 & 15 & 14 & 21 & 17 & 5 & 10 & 10 & 5 & 17 & 21 & 14 & 15 & 19 & 20 & 11 & 7 & 22 \\
0 & 22 & 15 & 19 & 5 & 13 & 14 & 2 & 17 & 7 & 12 & 3 & 20 & 11 & 16 & 6 & 21 & 9 & 10 & 18 & 4 & 8 & 1 \\
0 & 22 & 19 & 14 & 7 & 21 & 10 & 20 & 5 & 11 & 15 & 17 & 17 & 15 & 11 & 5 & 20 & 10 & 21 & 7 & 14 & 19 & 22 \\
0 & 22 & 21 & 20 & 19 & 18 & 17 & 16 & 15 & 14 & 13 & 12 & 11 & 10 & 9 & 8 & 7 & 6 & 5 & 4 & 3 & 2 & 1 \\
22 & 22 & 22 & 22 & 22 & 22 & 22 & 22 & 22 & 22 & 22 & 22 & 22 & 22 & 22 & 22 & 22 & 22 & 22 & 22 & 22 & 22 & 22
\end{pmatrix}$$

$k = 29$:

$$P = \begin{pmatrix}
1 & 0 & 0 & 0 & 0 & 0 & 0 & 0 & 0 & 0 & 0 & 0 & 0 & 0 & 0 & 0 & 0 & 0 & 0 & 0 & 0 & 0 & 0 & 0 & 0 & 0 & 0 & 0 & 0 \\
0 & 28 & 14 & 19 & 7 & 23 & 24 & 4 & 18 & 16 & 26 & 21 & 12 & 20 & 2 & 27 & 9 & 17 & 8 & 3 & 13 & 11 & 25 & 5 & 6 & 22 & 10 & 15 & 1 \\
0 & 28 & 7 & 16 & 9 & 22 & 4 & 13 & 24 & 5 & 20 & 23 & 1 & 6 & 25 & 25 & 6 & 1 & 23 & 20 & 5 & 24 & 13 & 4 & 22 & 9 & 16 & 7 & 28 \\
0 & 28 & 18 & 15 & 24 & 16 & 20 & 6 & 3 & 7 & 2 & 10 & 17 & 25 & 8 & 21 & 4 & 12 & 19 & 27 & 22 & 26 & 23 & 9 & 13 & 5 & 14 & 11 & 1 \\
0 & 28 & 9 & 5 & 6 & 9 & 13 & 5 & 4 & 4 & 6 & 22 & 28 & 22 & 13 & 13 & 22 & 28 & 22 & 6 & 4 & 4 & 5 & 13 & 9 & 6 & 5 & 9 & 28 \\
0 & 28 & 19 & 21 & 16 & 25 & 7 & 9 & 15 & 23 & 18 & 2 & 12 & 24 & 3 & 26 & 5 & 17 & 27 & 11 & 6 & 14 & 20 & 22 & 4 & 13 & 8 & 10 & 1 \\
0 & 28 & 24 & 7 & 4 & 5 & 6 & 22 & 20 & 9 & 25 & 16 & 1 & 13 & 23 & 23 & 13 & 1 & 16 & 25 & 9 & 20 & 22 & 6 & 5 & 4 & 7 & 24 & 28 \\
0 & 28 & 12 & 12 & 1 & 1 & 1 & 28 & 17 & 1 & 17 & 12 & 17 & 1 & 12 & 17 & 28 & 12 & 17 & 12 & 28 & 12 & 1 & 28 & 28 & 28 & 17 & 17 & 1 \\
0 & 28 & 6 & 4 & 22 & 6 & 5 & 4 & 13 & 13 & 22 & 9 & 28 & 9 & 5 & 5 & 9 & 28 & 9 & 22 & 13 & 13 & 4 & 5 & 6 & 22 & 4 & 6 & 28 \\
0 & 28 & 3 & 11 & 20 & 7 & 25 & 13 & 27 & 24 & 8 & 14 & 12 & 23 & 19 & 10 & 6 & 17 & 15 & 21 & 5 & 2 & 16 & 4 & 22 & 9 & 18 & 26 & 1 \\
0 & 28 & 16 & 23 & 5 & 13 & 9 & 6 & 7 & 22 & 21 & 25 & 1 & 4 & 20 & 20 & 4 & 1 & 25 & 24 & 22 & 7 & 6 & 9 & 13 & 5 & 23 & 16 & 28 \\
0 & 28 & 8 & 27 & 23 & 20 & 16 & 5 & 19 & 25 & 14 & 26 & 17 & 7 & 18 & 11 & 22 & 12 & 3 & 15 & 4 & 10 & 24 & 13 & 9 & 6 & 2 & 21 & 1 \\
0 & 28 & 4 & 9 & 13 & 4 & 22 & 9 & 6 & 6 & 13 & 5 & 28 & 5 & 22 & 22 & 5 & 28 & 5 & 13 & 6 & 6 & 9 & 22 & 4 & 13 & 9 & 4 & 28 \\
0 & 28 & 2 & 3 & 25 & 24 & 23 & 22 & 8 & 20 & 10 & 11 & 12 & 16 & 14 & 15 & 13 & 17 & 18 & 19 & 9 & 21 & 7 & 6 & 5 & 4 & 26 & 27 & 1 \\
0 & 28 & 1 & 1 & 28 & 28 & 28 & 28 & 1 & 28 & 1 & 1 & 1 & 28 & 1 & 1 & 28 & 1 & 1 & 1 & 28 & 1 & 28 & 28 & 28 & 28 & 1 & 1 & 28 \\
0 & 28 & 15 & 10 & 7 & 23 & 24 & 4 & 11 & 16 & 3 & 8 & 17 & 20 & 27 & 2 & 9 & 12 & 21 & 26 & 13 & 18 & 25 & 5 & 6 & 22 & 19 & 14 & 1 \\
0 & 28 & 22 & 13 & 9 & 22 & 4 & 13 & 5 & 5 & 9 & 6 & 28 & 6 & 4 & 4 & 6 & 28 & 6 & 9 & 5 & 5 & 13 & 4 & 22 & 9 & 13 & 22 & 28 \\
0 & 28 & 11 & 14 & 24 & 16 & 20 & 6 & 26 & 7 & 27 & 19 & 12 & 25 & 21 & 8 & 4 & 17 & 10 & 2 & 22 & 3 & 23 & 9 & 13 & 5 & 15 & 18 & 1 \\
0 & 28 & 20 & 24 & 6 & 9 & 13 & 5 & 25 & 4 & 23 & 7 & 1 & 22 & 16 & 16 & 22 & 1 & 7 & 23 & 4 & 25 & 5 & 13 & 9 & 6 & 24 & 20 & 28 \\
0 & 28 & 10 & 8 & 16 & 25 & 7 & 9 & 14 & 23 & 11 & 27 & 17 & 24 & 26 & 3 & 5 & 12 & 2 & 18 & 6 & 15 & 20 & 22 & 4 & 13 & 21 & 19 & 1 \\
0 & 28 & 5 & 22 & 4 & 5 & 6 & 22 & 9 & 9 & 4 & 13 & 28 & 13 & 6 & 6 & 13 & 28 & 13 & 4 & 9 & 9 & 22 & 6 & 5 & 4 & 22 & 5 & 28 \\
0 & 28 & 17 & 17 & 1 & 1 & 1 & 28 & 12 & 1 & 12 & 17 & 12 & 1 & 17 & 12 & 28 & 17 & 12 & 17 & 28 & 17 & 1 & 28 & 28 & 28 & 12 & 12 & 1 \\
0 & 28 & 23 & 25 & 22 & 6 & 5 & 4 & 16 & 13 & 7 & 20 & 1 & 9 & 24 & 24 & 9 & 1 & 20 & 7 & 13 & 16 & 4 & 5 & 6 & 22 & 25 & 23 & 28 \\
0 & 28 & 26 & 18 & 20 & 7 & 25 & 13 & 2 & 24 & 21 & 15 & 17 & 23 & 10 & 19 & 6 & 12 & 14 & 8 & 5 & 27 & 16 & 4 & 22 & 9 & 11 & 3 & 1 \\
0 & 28 & 13 & 6 & 5 & 13 & 9 & 6 & 22 & 22 & 5 & 4 & 28 & 4 & 9 & 9 & 4 & 28 & 4 & 5 & 22 & 22 & 6 & 9 & 13 & 5 & 6 & 13 & 28 \\
0 & 28 & 21 & 2 & 23 & 20 & 16 & 5 & 10 & 25 & 15 & 3 & 12 & 7 & 11 & 18 & 22 & 17 & 26 & 14 & 4 & 19 & 24 & 13 & 9 & 6 & 27 & 8 & 1 \\
0 & 28 & 25 & 20 & 13 & 4 & 22 & 9 & 23 & 6 & 16 & 24 & 1 & 5 & 7 & 7 & 5 & 1 & 24 & 16 & 6 & 23 & 9 & 22 & 4 & 13 & 20 & 25 & 28 \\
0 & 28 & 27 & 26 & 25 & 24 & 23 & 22 & 21 & 20 & 19 & 18 & 17 & 16 & 15 & 14 & 13 & 12 & 11 & 10 & 9 & 8 & 7 & 6 & 5 & 4 & 3 & 2 & 1 \\
28 & 28 & 28 & 28 & 28 & 28 & 28 & 28 & 28 & 28 & 28 & 28 & 28 & 28 & 28 & 28 & 28 & 28 & 28 & 28 & 28 & 28 & 28 & 28 & 28 & 28 & 28 & 28 & 28
\end{pmatrix}$$